\documentclass[fleqn,10pt]{olplainarticle}

\usepackage[capitalise]{cleveref}

\usepackage[super]{nth}

\usepackage{url}

\title{Early Career Wins and Tournament Prestige Characterize Tennis Players' Trajectories}

\author[1,2,3,4,*]{Chiara Zappal\`a}
\author[3,4,$\dag$]{Sandro Sousa}
\author[3,$\dag$]{Tiago Cunha}
\author[2]{Alessandro Pluchino}
\author[2,5]{Andrea Rapisarda}
\author[4,3,5,6]{Roberta Sinatra}
\affil[1]{Center for Collective Learning, Corvinus Institute for Advanced Studies (CIAS), Corvinus University, 1093 Budapest, Hungary}
\affil[2]{Department of Physics and Astronomy, University of Catania and INFN sezione di Catania, 95123 Catania, Italy}
\affil[3]{NEtwoRks, Data, and Society (NERDS), Computer Science Department, IT University of Copenhagen, 2300 Copenhagen, Denmark}
\affil[4]{Center for Social Data Science (SODAS), University of Copenhagen, 1353 Copenhagen, Denmark}
\affil[5]{Complexity Science Hub, 1080 Vienna, Austria}
\affil[6]{ISI Foundation, 10126 Turin, Italy}
\affil[$^*$]{Corresponding author. E-mail: chiara.zappala@uni-corvinus.hu}
\affil[$\dag$]{These authors contributed equally to this work.}
\keywords{Network Science applications, Success, Sports analytics}

\begin{abstract}
Success in sports is a complex phenomenon that has only garnered limited research attention.
In particular, we lack a deep scientific understanding of success in sports like tennis and the factors that contribute to it.
Here, we study the unfolding of tennis players' careers to understand the role of early career stages and the impact of specific tournaments on players' trajectories.
We employ a comprehensive approach combining network science and analysis of ATP tournament data and introduce a novel method to quantify tournament prestige based on the eigenvector centrality of the co-attendance network of tournaments.
Focusing on the interplay between participation in central tournaments and players' performance, we find that the level of the tournament where players achieve their first win is associated with becoming a top player.
This work sheds light on the critical role of the initial stages in the progression of players' careers, offering valuable insights into the dynamics of success in tennis.
\end{abstract}

\begin{document}

\flushbottom
\maketitle
\thispagestyle{empty}

\section*{Introduction}
Understanding the complex mechanisms behind the origin of success is a challenging task that has captured the attention of researchers in the past few years, as it encompasses a wide range of systems.
To mention some examples, paper citations \cite{Sinatra2016,Clauset2017,Fortunato2018} and funding \cite{bol2018matthew} in science, start-ups \cite{bonaventura_predicting_2020}, show business \cite{Williams2019}, art \cite{Fraiberger2018} and cryptoart \cite{nadini2021mapping,vasan2022quantifying} ecosystems, music \cite{Salganik2006,janosov_elites_2020,kang2022analyzing}, and other creative careers \cite{wang_success_2019,Janosov2020}, have been investigated to date.

A common issue in these systems is to unambiguously distinguish between performance and success \cite{murray_human_2003}.
Whereas performance refers to objectively measurable accomplishments in a particular field \cite{Yucesoy2016}, such as the publication record of a scientist \cite{lehmann2006measures}, success represents the reward of a given level of performance \cite{Barabasi2018}, intended as its collective recognition, such as the citation impact in science \cite{Barabasi2018,Sinatra2016} or prize and awards in fields like music \cite{janosov_elites_2020}.
However, assessing the impact of creative work only through prizes and fame might fail to consider the abilities of the individuals involved, that is, to disentangle performance from success \cite{murray_human_2003}.

Sports allow us to overcome this issue.
First, they offer objective metrics for evaluating performance, like the winning record of an athlete or a team \cite{Radicchi2012}.
Most importantly, successful players are identifiable by the reward system of the sport itself, i.e. rankings based on score systems, especially in individual sports based on knockout tournaments.
In fact, sports rankings depend on criteria that are external to the athletes' performance, i.e., the quality of a tournament (also called tourney) has an \textit{apriori} fixed value and points are distributed accordingly to the round reached in it \cite{Radicchi2011a}.
Therefore, unlike the previous literature on sports \cite{Yucesoy2016}, here we consider the ranking as a metric of success that is inherently provided by sports rules, not determined by the popularity of players.

Only a few works have analyzed sports disciplines from a complex systems perspective \cite{Yucesoy2016,tennant_complexity_2017,pappalardo2018quantifying,Sobkowicz2020,Zappala2022}.
Additionally, we lack a systematic analysis of the key factors behind the unfolding of sports careers.
Specifically, the impact of early career stages on the future achievements of players has rarely been taken into account,  despite the proven importance of these initial stages in many different kinds of careers \cite{Petersen2011}.

The determinants of successful careers in sports remain elusive.
Often, we imagine the top players as predestined champions who need to be extraordinarily talented and hard-working to get to the top \cite{zappala_paradox_2023}.
Yet, evidence suggests that a combination of talent and effort does not guarantee success \cite{mauboussin2012success,Frank2016}.
Rather, some initial fortuitous events might play a role in shaping the evolution of top players' careers, as shown for individual sports \cite{Sobkowicz2020,Zappala2022,zappala_paradox_2023}.
The role of chance at the early stages can be later amplified by a cumulative advantage dynamic \cite{perc2014matthew}.
The aforementioned factors provide compelling reasons to delve into the trajectories of players' careers, i.e., the temporal sequences of the competitions they attended.

Here, we focus on tennis, for which a limited number of studies have examined the careers of top players \cite{Yucesoy2016,Radicchi2011a,zappala_paradox_2023}, yet little is known about the determinants of their future success at the beginning of their career.
In particular, we aim to comprehend the role of early access to prestigious tournaments in shaping the future of top tennis players.
Specifically, we analyze the career progressions of professional male players between 2000 and 2019.
We collect data from the official rankings of the Association of Tennis Professionals (ATP) \cite{ATP}, along with the results of matches from various tournaments \cite{ATPgit}.
The top tennis athletes are identified by their career peak, which is determined by the highest number of ranking points they have achieved in the ATP rankings.
Through our analysis, we observe distinctive characteristics among accomplished players compared to others, including longer career spans and a pattern of consistently higher ranking points throughout their career's initial stages.

Our hypothesis is that the emergence of top players is associated with their participation in high-level competitions during the early stages of their career.
This effect is analogous to the results on high-profile artists whose reputation is associated with the exhibition of their initial work at prestigious institutions \cite{Fraiberger2018}.
To test this, we introduce a novel approach to quantify the level of ATP tournaments, which not only includes their historical prestige but also takes into account the participation of players.
This method, based on network science principles, presents a contribution that, to our knowledge, has not been explored in the existing literature on tennis.
Previous studies have used networks solely to analyze match relationships \cite{Radicchi2011a,Breznik2015,Aparicio2016,zappala_paradox_2023}.
We expand upon this by constructing a network of co-attendance among tennis tournaments, where nodes represent tourneys, and links are created based on players' trajectories, that is, a link connects two tourneys if there is at least one player who competed in these two tourneys during his career.
Consequently, we can establish connections between competitions that may be geographically distant or temporally separated.
By leveraging this co-attendance network, we derive a measure of tournament prestige using eigenvector centrality \cite{bonacich1987power}, following the methodology of Ref.~\cite{Fraiberger2018}.

We will show that the level of the tourney where players secure their first match win allows us to characterize future successful players.
In fact, top players tend to win for the first time in highly central tournaments, even though the average level of the first tournaments they attend is not different from that of less accomplished players.

The outline of this paper is the following.
First, we analyze the differences between top players and their counterparts, focusing on the initial stages of their careers.
Second, we build the co-attendance network to identify highly central tournaments.
Third, we link the tourney level with players' performance, assessed by their first match win, and we find an association between this link and becoming top players.
Finally, we check the robustness of our findings using two distinct null models.

\section*{Characterizing patterns in tennis careers}
We study the evolution of the careers of male professional tennis players from 2000 to 2019.
We obtained data from the official ranking of the ATP \cite{ATP,ATPgit}, together with the match results of different tournaments: Grand Slam (the competitions with the highest value in terms of winner points), Masters 1000, ATP 500 and 250, Challenger (the competitions with the lowest value in our dataset) \cite{ATPgit}.
We selected players who started their careers in the timespan of our dataset and had at least two years of activity.
We consider 3,459 players and 651 tourneys, specifically 4 Grand Slams, 11 Masters 1000, 98 ATP 500 and 250, and 538 Challengers.

To distinguish between top and less successful tennis players, we group them according to the maximum amount of points they reached in the ATP ranking, which ranks players based on the score points they accumulate during a season \cite{ATP}.
We can conceive the ATP ranking as a first proxy of success, as it might weigh similar outcomes of players' performance, which would be winning or losing one or more matches, in very different ways.
For instance, winning a match in the round of 32 awards 5 points in a Challenger and 90 points in a Grand Slam.
Thus, rather than relying on popularity to quantify success in sports \cite{Yucesoy2016}, we explore the dynamics of success embedded in tennis rules, neglecting the influence of subsequent elements such as prize money, income, popularity, sponsors, etc.
Moreover, we use the highest number of points players reach in the ATP ranking (namely, their career peak) instead of ranking placements.
The reason is that the point totals of players with consecutive ranks can vary significantly.
For example, consider three players ranked 1, 2, and 3, with point totals of 12,000, 10,000, and 9,995, respectively.
Although players ranked 1 and 2 are only one position apart from each other, as well as players ranked 2 and 3, there is a greater point difference (2,000 points) between players ranked 1 and 2 than between players ranked 2 and 3 (5 points).
Therefore, using point totals instead of placements allows us to assess differences between players more accurately.
Also, it lets us compare rankings with varying numbers of players over the years.

\begin{figure}[p]
\centering
\includegraphics[width=\textwidth]{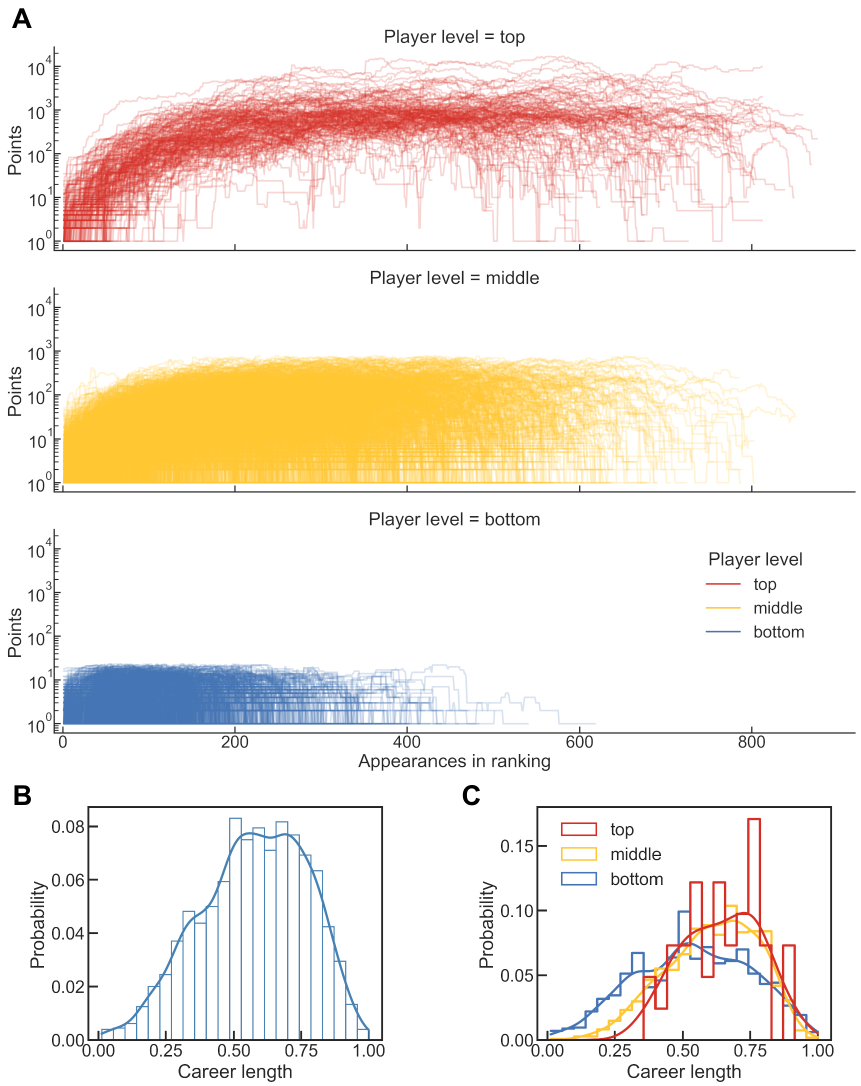}
\caption{Group splitting among the players, based on their career peak.
\textbf{A} Timelines of the ranking points accumulated through the career of players within each group.
All individual sequences have been aligned according to appearances in ATP tournaments.
Differences between player groups are visible in both (i) accumulated points and (ii) the number of appearances in the ranking.
\textbf{B} Career peak distribution for the whole community of players relative to the length of the player's career.
The peak occurs after the first half of their careers.
\textbf{C} Career peak distributions after splitting players into groups.
For all groups, the ranking peak appears between 50\% and 75\% of the players' career length.
Histograms in panels B-C are normalized so that bar heights sum to 1 and are reported with a Kernel Density Estimation of the data (continuous curves).}
\label{fig:1}
\end{figure}

We divide male tennis players into three groups: We identify top players as those with a career peak above the \nth{95} percentile (top $5\%$); bottom players are within the \nth{40} percentile (bottom $40\%$); the $55 \%$ left falls in the middle.
Panel A of \cref{fig:1} reports the evolution of players' ranking points, grouped by their career peak, showing all individual timelines.
Note that players can start their careers at different times, so we aligned their sequences using the number of updates in the ATP rankings.
Specifically, we consider updates that occur when we have a data point, that is when a player attends a given tournament in our dataset.

Our analysis is based on the peak of professional tennis players' careers, prompting the question of whether this peak is obtained at a consistent time across all individuals within our dataset.
To answer this question, we look at the time distribution of career peak, first considering the aggregated data, then each group separately (respectively, panels B and C of \cref{fig:1}).
To avoid right-censoring bias \cite{wu1988estimation}, we exclude active players from \cref{fig:1}B-C (more details are provided in the SI, where we report the case without the right-censoring correction in Fig.~S1).
To deal with different career lengths, we normalize the time of the career peak of each player according to their career duration.
We find a common tendency for the peak to occur after the first half of players' careers in all three groups.
This result, previously observed only for the top players \cite{Schulz1988,Guillaume2011}, suggests that peak time is not closely related to individual success.

\begin{figure}[h!]
\centering
\includegraphics[scale=1]{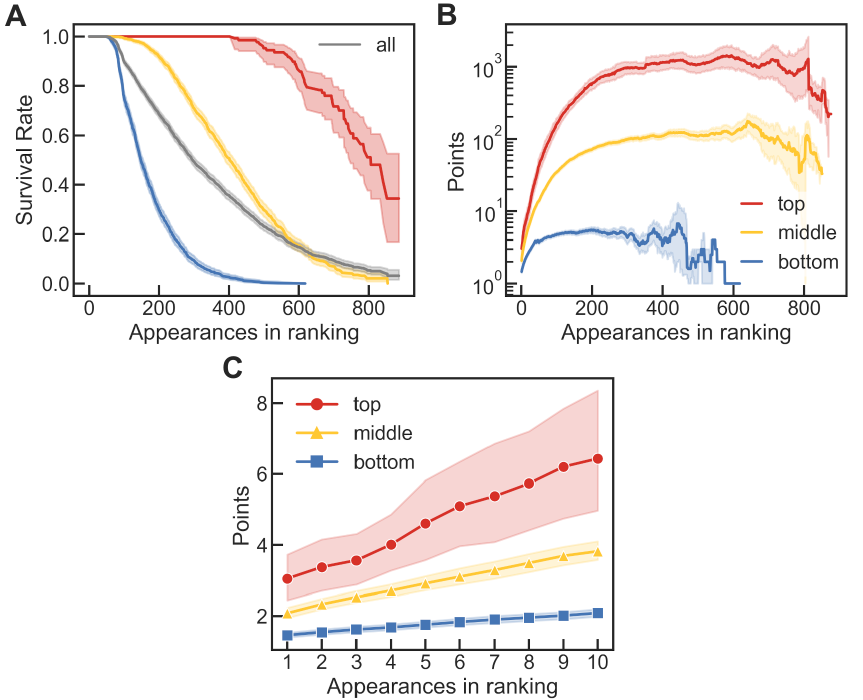}
\caption{Overall trend of male tennis players in ATP ranking.
\textbf{A} Survival rate of players in the ranking from 2000 to 2019.
Top players have longer careers (red curve) compared to middle (yellow) and bottom (blue) players.
\textbf{B} Average evolution of ranking points for the groups of players.
Top players (in red) consistently accumulate more points compared to the others.
\textbf{C} Focus on players' first ten appearances of their careers, which shows a clear group separation between top (red dots), middle (yellow triangles) and bottom (blue squares) players.
}
\label{fig:2}
\end{figure}

Observing the individual sequences of the three groups in \cref{fig:1}A, we notice marked differences between them, both in ranking appearances and accumulated points.
In particular, the bottom players have shorter careers compared to the other groups.
We further investigate this by looking at the survival rate \cite{Davidson-Pilon2019} of tennis players in our dataset, bearing in mind that in this context ``surviving'' at time $t$ means still playing, or in other words, being in the ATP ranking.
The results are shown in panel A of \cref{fig:2}: The bottom players' curve (in blue) goes rapidly to zero; conversely, the top players' curve (in red) decays more slowly, meaning that they have longer professional careers, in line with previous works \cite{Baker2013}.
Finally, the middle players' survival curve (in yellow) lies between the other two.
For completeness, we also report the survival rate of all players (in gray).

To highlight when the group differences in accumulated points appear, we take the average of the sequences shown in \cref{fig:1}A, which results in the trend of \cref{fig:2}B: We can observe that the top players have more ranking points compared to the others.
Such a discrepancy in the number of points could be interpreted as a mere artifact of our definition of top/middle/bottom players.
Yet, this difference emerges from the beginning, as indicated in panel C of \cref{fig:2}, which zooms in on the points cumulated in only the first ten tournaments of a player's career.
Note that here we include the careers of all players, thus including active players.
See Fig.~S2 of the SI for an analysis that considers only those players who started and ended their careers in the dataset.

\begin{figure}[h!]
\centering
\includegraphics[width=\textwidth]{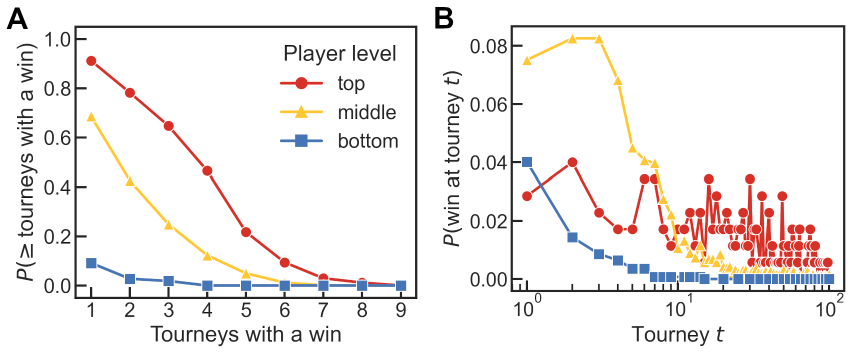}
\caption{Performance comparison for tennis players, considering their first tourneys with a win.
\textbf{A} Probability for players of having at least a given number of distinct tourneys in which they have won one or more matches.
Top players (red dots) have higher chances of winning more matches, once they have won their first, within the first ten tournaments they attended.
\textbf{B} Probability for players of winning their first match in a tourney $t$, with $1 \leq t \leq 100$.
Top players (red dots) win their first match later than the others.
}
\label{fig:3}
\end{figure}

The initial gap in the average amount of points between the top players and the others may arise from different mechanisms.
A first explanation for such a gap may lie in the differences in players' performance.
That is, top players may win more matches from the early stages of their careers, leading to the gap forming.
To compare performance across the groups, we first consider the number of competitions in which a player wins at least one match.
Panel A of \cref{fig:3} shows the probability $P$ that a player, with at least one match won, reports a win in more than a given number of tourneys, within the first ten (see Methods for the mathematical definition).
Top players (red dots) have higher chances of winning more matches, once they have won their first, at the beginning of their career.
However, if we look at the probability $P$ of players winning their first match in a certain tourney $t$ after they turned professional (\cref{fig:3}B, see Methods for the mathematical formulation), we do not see top players emerge.
On the contrary, top players tend to win their first match later than the others.

The results of \cref{fig:3} show that, even though top players achieve more victories after their initial one, they have difficulty in winning their very first match during the early stages of their career.
This counterintuitive behavior points out that players' performance is not enough to explain the formation of the gap between top and less accomplished athletes.
Hence, other mechanisms might be at play.
For instance, our analysis so far has not taken into account the prestige of the different tournaments that players can attend at the beginning of their careers.
Therefore, having illustrated the scenario of the initial stages of players' careers in men's professional tennis, we investigate the influence of the first tournaments they can access, together with their results.
We aim to untangle the importance of early participation in more prestigious competitions from how players perform in those competitions.

\section*{The co-attendance network}
\begin{figure}[h!]
\centering
\includegraphics[width=\textwidth]{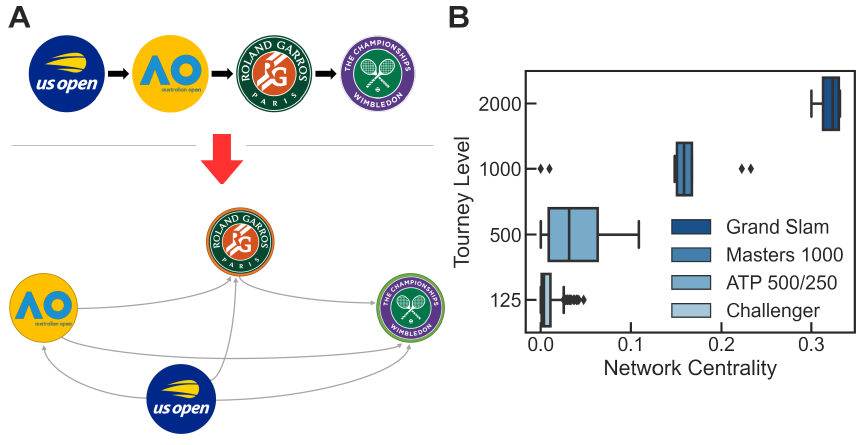}
\caption{Network diagram and comparison between network centrality and ATP level of tournaments.\\
\textbf{A} Example of a co-attendance network generated by the trajectory of a player's career.
In this case, a player's trajectory consists of four Grand Slam tourneys, from the US Open to Wimbledon (sequence on top of the panel).
The resulting network has four nodes and six links.
\textbf{B} Comparison between network-based prestige, expressed by the eigenvector centrality ($x$ axis), and the hierarchy of tourneys in the ATP, represented by the points assigned to the winner ($y$ axis).
The plot shows that the network-based prestige is positively correlated with the ATP hierarchy.}
\label{fig:4}
\end{figure}

Early access to prestigious tournaments could affect the career trajectories of players in the ATP circuit in a non-trivial way.
Those trajectories then create complex relationships between players and tourneys.
Indeed, players do not have the possibility to participate in all available tournaments and choose which tourney to sign up for based not only on their own skills but also on the characteristics of tourneys themselves (e.g., the court), their relevance during a season (preceding or succeeding famous events, for example), and their prestige.
One could quantify tournament prestige from their prizes in terms of ranking points.
Yet, assessing the tourney level based only on prizes does not capture the prestige perceived by the players and determined by their choices.
Following Ref.~\cite{Fraiberger2018}, we propose a new method to assess the level of a certain ATP tournament, which not only captures its historical prestige but also includes the reciprocal influence of players and the reputation of a given competition.
We define a network where nodes are ATP tourneys and links are determined by players' careers.
A directed link $\left(i, j \right)$ is created when a player first competes in tourney $i$, then in tourney $j$, and is weighted by the number of players who have the same competition trajectory \cite{Fraiberger2018} (see panel A of \cref{fig:4} for an example).
Note that we consider the effects of all the possible competition trajectories that players made along their careers in the data.
In this way, the movements of players link competitions far in space and time, and recurrent movements signal that those competitions tend to co-occur in players' careers.

The resulting network has 659 nodes and 255,055 edges.
We focus on the largest strongly connected component of the original network, which has 651 nodes and 254,583 links; from now on, we refer to the largest component as our network (see Table~S1 in the SI for more details on the features of the network, such as its density and clustering coefficient).

From the co-attendance network, we can extract a measure of tourney prestige that correlates with the importance of tournaments in terms of their points.
The prestige of a tournament can be derived from the topology of the network, using the eigenvector centrality \cite{newman2018networks,Fraiberger2018} (see also Methods for the mathematical definition we used).
This definition captures the historical level of the competitions (\cref{fig:4}B), expressed by the maximum amount of points assigned to each tournament category (the allocation of points rewarded per tournament is explained in the Methods section and summarized in Table~S2 of the SI).

Based on their centrality, we divide competitions into three groups: The most prestigious tournaments are in the top $10\%$ (above the \nth{90} percentile), and we refer to them as high-level tourneys; the bottom $50\%$ (below the \nth{50} percentile) of the competitions are labeled as low-level tournaments; the others fall into the medium-level group of events.
Having defined the level of the tourneys, we now focus on the possible connections between those levels and the success of players in the ATP circuit.
These connections are crucial to determine whether the opportunity of competing in a given tournament could be more relevant than, or at least as relevant as, players' abilities.

\section*{Early access to prestigious tournaments and the impact of the first win}
\begin{figure}[p]
\centering
\includegraphics[width=.98\textwidth]{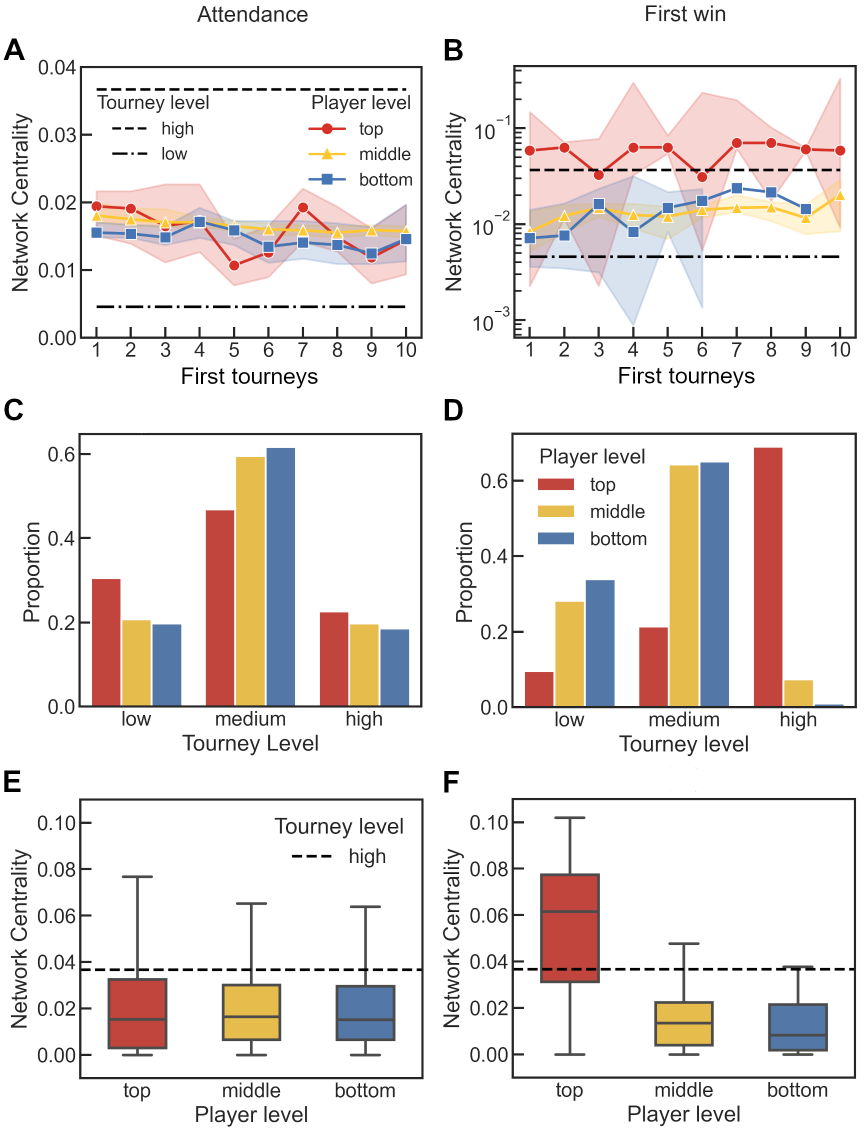}
\caption{Analysis of players' attendance (left panels) and first win (right panels).
\textbf{A} Centrality trend for the first ten tournaments attended per group of players.
\textbf{B} Centrality trend of the first match win within the first ten tourneys per group of players.
\textbf{C} Fractions of players who participated in a tournament of a given level within the first ten competitions.
Bars of the same color, each identifying a given group of players, add to 1.
\textbf{D} Fractions of players who have their first win in a tournament of a given level within their first ten competitions.
Again, bars of the same color sum up to 1.
\textbf{E} Graphical representation of the distribution of the centrality of the first ten tournaments for each group of players.
\textbf{F} Graphical representation of the distribution of the centrality of the first win for each group of players.
The dashed (dashed-dotted) lines refer to the high (low) level threshold of tourney splitting.
The plots show that top players (in red) behave differently from the others only when considering their first match win.
}
\label{fig:5}
\end{figure}

We check the possible differences in tournament attendance within the first ten tournaments of athletes' professional careers in the ATP circuit (left panels of \cref{fig:5}).
First, in \cref{fig:5}A, we observe the eigenvector centrality of the first ten tournaments for each group of players based on their career peak.
We find that, at the beginning of their career, players attend competitions with comparable centrality, having median values in between the thresholds (dashed lines) of tournament splitting (we consider the median due to the asymmetric distribution of the centrality in our network, shown in Fig.~S3 of the SI).
Only after a considerable number of tourneys do top players attend only high-level tournaments, which means that they consistently participate in events having a central position in the co-attendance network (see Fig.~S4 in the SI).
Then, we inspect the fraction of players who enter a certain tourney of a given level at the beginning of their career (panel C of \cref{fig:5}).
We do not observe a pronounced prevalence of future top players in high-level competitions (red bars in \cref{fig:5}C).

Interestingly, we find no significant differences in the prestige of tournaments players can access when their careers start.
One might argue that the seasonality of tournaments plays a role, hence affecting the centrality of players' first competitions: In other words, if the centrality of the first tourneys of the season is around the median value, we should expect the trend observed in panel A of \cref{fig:5}.
Nonetheless, professional players can start their career on the ATP tournament circuit at any time during a competitive season, coincident with the calendar year.
Therefore, the centrality of the opening tournaments of the season (in other words, the tournaments organized in January/February) does not determine the entire trend of \cref{fig:5}A.
Consequently, players' first attended tournaments can vary widely from athlete to athlete.
It is also worth mentioning that we neglect the influence of the junior circuit on players' professional development.
According to some studies \cite{Reid2007,Kovalchik2017,Pingwei2018}, the youth career could impact the future success of an athlete in tennis.
Even if that impact were not a prerequisite for professional success \cite{Guillaume2011,Brouwers2012}, young players who performed well at the junior level could be favored to access more prestigious ATP venues.
However, such an effect, if present, does not create a significant gap among players in terms of the level of the first attended tourneys.
We specify that we do not differentiate players by age or other factors like country of origin or physical characteristics (e.g., height, left-handed or right-handed, etc.).

Since no patterns emerge when looking at tournament attendance, one might ask if there are differences related to performance.
In our framework, tennis performance is expressed by the outcome of matches.
Thus, we check for patterns linking the victory of matches and the start of players' professional careers in the ATP circuit.
To do so, we focus on the first win of a match at the beginning of tennis players' careers.
Specifically, we are interested in the first victory in the main draw (i.e., the starting lineup of a tourney after the qualification rounds) of the first ten tournaments they attended.
We consider the first match win because it allows us to directly compare the outcome of players' performance for all the competitors.
To visualize the relationship between the first win and the tourney level, in terms of centrality, we refer to the right panels of \cref{fig:5}.
In \cref{fig:5}B, the eigenvector centrality of the top players is the only one above the threshold of high-level competitions.
We find that most of the top players (around $70\%$) have their first win in the main draw of a high-level tournament (\cref{fig:5}D).
Furthermore, only top players can be identified by looking at the prestige of their first win in a match.
Both middle and bottom players have similar behavior (\cref{fig:5}D), and their first victory rarely occurs in high-level competitions.

To better understand the discrepancy in the behavior of top players, either when we consider only their attendance or when we add their first win, we compare the boxplots of the centrality of the players' first tournaments with that of their first win (see panels E and F of \cref{fig:5}, respectively, while a fine-grained visualization is available in the SI, Fig.~S5).
In this way, we observe a clear difference between the two situations.
In \cref{fig:5}E, there is no distinction between the top, middle, and bottom players, with respect to the network-based prestige of the first tournaments they attended.
Panel F of \cref{fig:5}, instead, shows that the average centrality of the first win of the top players crosses the threshold of high-level tourneys.
In particular, the top players' boxplot is the only one that changes from panel E to F, which means that the higher prestige of the top players' first win cannot be explained by the average level of the first attended tourneys.
Panels E and F of \cref{fig:5} highlight the relationship between the prestige of the tournament in which the top players win their first match in a main draw and their future success.
To further validate our analysis, we computed the correlation between (1) the player's career peak and the median centrality of the first tournaments they played; and (2) the player's career peak and the centrality of the tourney where they got their first win in a main draw.
We use Spearman's correlation coefficient and find that $r_{s,1} = -0.008^{\:\mathrm{ns}}$ and $r_{s,2} = 0.47^{\:***}$, where $^{\:\mathrm{ns}}$ and $^{\:***}$ indicate the level of significance of the two values, that is, the p-value is greater than 0.05 (not significant) and less than 0.001 (significant), respectively.

Whether comparing players in groups or directly through their maximum number of points, we conclude that the prestige of the tournament where they first win a match in the main draw is a revealing factor for the future career of male tennis players.
It should be noted that taking the qualification rounds into account does not appreciably change our findings (Fig.~S6 of the SI). 
Furthermore, we do not assume that players should attend at least ten tournaments to be in the dataset, and we do not exclude active players, but adding these constraints does not significantly alter our results (see Figs.~S7-S8 in the SI).

\section*{Significance of the results}
\begin{figure}[p]
\centering
\includegraphics[width=.98\textwidth]{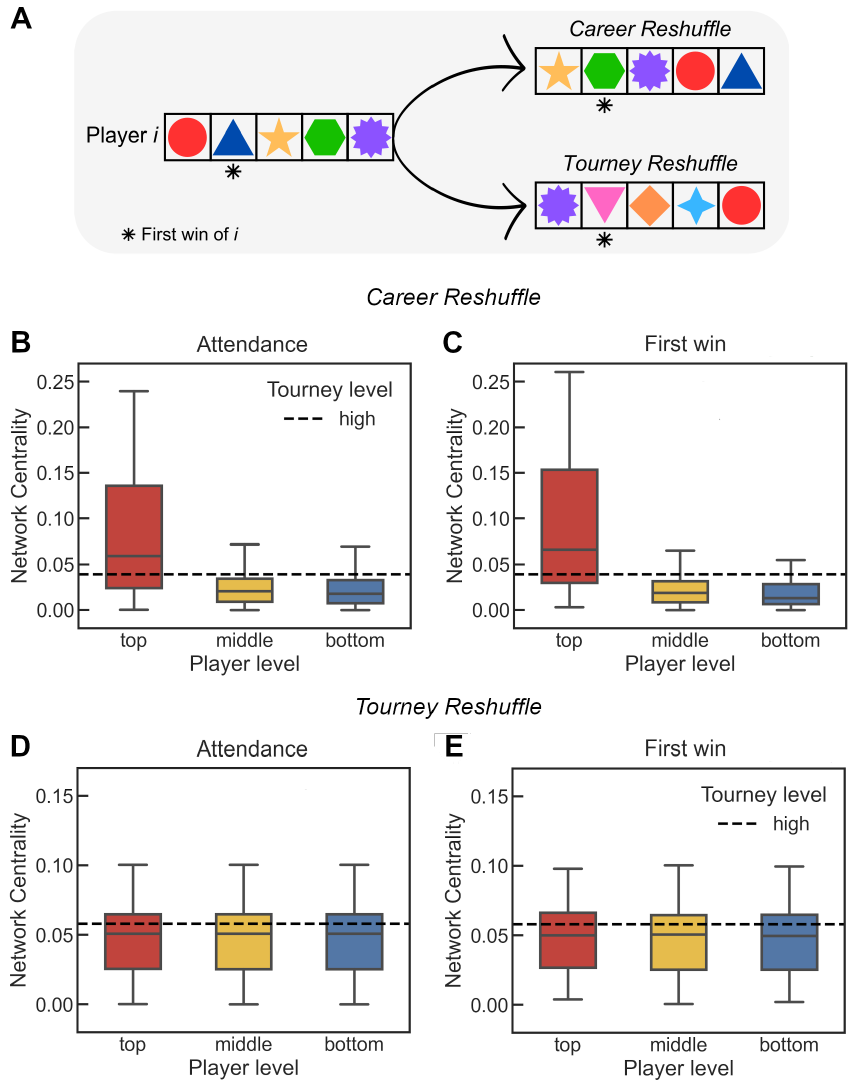}
\caption{Results of the two distinct null models.
\textbf{A} Comparison of the different randomizations: in the first model (career reshuffle), player $i$ has the same tourneys, but the temporal sequence is different (same shapes in a different order), while in the second model (tourney reshuffle) some competitions are different from the original sequence of tournaments of $i$ (some shapes have changed).
The asterisk marks the unchanged time of the first win (while the shapes, i.e., the tourneys, are different).
\textbf{B} Boxplots of the random distribution of the centrality of the first ten tournaments after players' careers have been reshuffled.
\textbf{C} Boxplots of the random distribution of the centrality of the first victory for each group of players after their careers have been reshuffled.
\textbf{D} Boxplots the random distribution of the centrality of the first ten tournaments, having randomized all the tournaments among the players.
\textbf{E} Boxplots of the random distribution of the centrality of the first win for each group of players, having randomized all tournaments among the players.
In both models, different from the data, the effect of attendance is indistinguishable from the first win.}
\label{fig:6}
\end{figure}

To assess the significance of our findings, we build two distinct null models for the network of co-attendance of ATP tournaments.
Building on previous work \cite{Sinatra2016,Fraiberger2018}, we proceed as follows (\cref{fig:6}A): In the first model, we reshuffle the careers of each player individually so that the events they play are the same but have a different temporal order; in the second model, we reshuffle all the competitions attended by the players, so that each player's career has the same number of events, but it can consist of different tourneys.
In both cases, all temporal correlations are destroyed.
We choose these two randomizations because they focus on different aspects: The first randomization preserves not only the number of competitions but also the actual events players attended; the second randomization preserves the number of tournaments of each player and the distribution of competitions among all players, destroying, however, the possible player-tourney correlations.

We repeat these two randomizations multiple times to create an ensemble of 500 random realizations.
We specify that in both configurations we preserve the information about the time of the first win as given by the data.
Therefore, the randomizations would only change the corresponding tournament in the sequence, but not \emph{when} the first win of a player occurred (as illustrated by the asterisk in \cref{fig:6}~A).
For each realization, we build the correspondent co-attendance network and evaluate tournament centrality, considering the prestige of competitions as done in the data.

We analyze the average distributions of tournament centrality per group of players, thus keeping the possible relationships between players' success and prestige of their initial ten competitions.
We follow the order of tourney attendance and define tournament levels based on their importance in the null models.
Panels B to E of \cref{fig:6} show that the null models cannot reproduce at the same time both the prestige of the tournaments attended and that of the first match win in the early stage of top players' professional career in tennis (\cref{fig:5}E-F).

The reshuffle of the individual sequences of tournaments per player increases the gap between top and middle-bottom players: given the cyclic nature of individual sports, where competitions repeat themselves every year around the same week, players are encouraged to attend the same tourneys season after season, to preserve or improve their amount of points.
It follows that reshuffling the careers of top players only anticipates those tournaments they start to play once they have already reached the top.
In contrast to empirical data, consequently, top players tend to compete more in high-level tournaments from the beginning of their professional careers, so that they are more likely to win their first match in highly central competitions (panels B-C of \cref{fig:6}).

On the other hand, the randomization of all tourneys destroys the cyclic trend of sports based on seasonal tournaments.
Thus, we do not expect significant differences in the level of competition among players or any eventual correlation between their career peak and their results.
Indeed, we observe in \cref{fig:6}D-E that there is no distinction between the groups and no patterns emerge in the prestige of their tournaments.

We also compare the mean value of Spearman's correlation coefficients for both null models over all configurations.
As described for the data, we computed the correlation between (1) the player's career peak and the centrality of the first tournaments they played, and (2) the player's career peak and the centrality of the tourney where they got their first win in a main draw.
The results of both randomizations are summarized in \cref{tab:1}.
When reshuffling individual sequences, we observe that the athlete's career peak is slightly positively correlated with both the centrality of the first tournaments attended and the centrality of the first win.
Instead, when we randomize all possible tournaments among the players, we find almost zero correlation in both cases.
\begin{table}[h!]
\centering
\begin{tabular}{lcc}
\hline
Null model & Attendance & First win \\
\hline
 & $\bar{r}_{s,1} \pm \sigma_{r_{s,1}}$ & $\bar{r}_{s,2} \pm \sigma_{r_{s,2}}$ \\
\cline{2-3}
Career reshuffle & $0.182 \pm 0.005$ & $0.14 \pm 0.03$ \\
Tourney reshuffle & $0.001 \pm 0.007$ & $0.01 \pm 0.03$ \\
Data & $-0.008^{\:\mathrm{ns}}$ & $0.47^{\:***}$ \\
\hline
\end{tabular}
\caption{Spearman's correlation coefficients for the null models, related to players' participation and first win, within the initial ten tournaments.
We also report the correlation coefficients found in the data for comparison.}
\label{tab:1}
\end{table}

Whereas we find a significant difference between these two correlation coefficients in the data, we observe that such a discrepancy is not significantly different from zero for both null models.
This means that the behavior we observe in the data cannot happen by chance, i.e., the discrepancy between the centrality of the first tourneys top players attend compared to where they first win a match in their professional career is not random.
Thus, the prestige of the tournament where male tennis players have their first win is a predictor of their future careers.

\section*{Discussion}
In this work, we analyze the career evolution of tennis players to uncover the key features that characterize top players and their future achievements.
To do so, we introduce a network-based ranking of tournaments that captures the underlying connections created by players' movements in the ATP circuit according to their attendance.
Our focus is on the early stages of tennis players' career and we look at the level of tourneys they attend upon entering the ATP circuit.
We find that participation in tournaments of different levels is not a good predictor of athletes’ success.
Instead, we find that the level of the tourney where players win their first match allows us to identify the top players.
We conclude that the first match win in highly central tournaments is a revealing factor for the future success of male tennis players.

We can speculate on possible explanations for this relationship between the prestige of the first win and the success in tennis.
Up-and-coming players who win at a central venue might have their visibility boosted, attracting the attention of the rest of the tennis community, especially talent scouts and tournament organizers.
The former could bring motivation, new staff, and perhaps even fans, ultimately reaching a broader audience through the media.
The latter could award promising players with a wild card, which would allow them to access more relevant tournaments without the required ranking (wild cards are awarded at the discretion of the organizers) \cite{ATP}.
These circumstances would boost players' confidence in the management of highly demanding matches, both physically and mentally.
Also, players with comparable performance, but in a less prestigious event, receive fewer ranking points.
Therefore, a first win in a prestigious competition paves the way for accessing more and more important tournaments.
Additionally, the economic benefits of winning in tennis (partially due to the prize money of the tourneys themselves, more commonly related to sponsorship and advertising) could play a role in shaping players' careers.

Our findings highlight the impact of the initial stages of players' careers, as a single match win can affect their future trajectories. 
Furthermore, they advocate for a deeper investigation of the economic implications that follow relevant sports results and might influence the professional development of players.

\section*{Methods}
\subsection*{Ranking point scale}
Professional male tennis players accumulate points in the ATP ranking during a season (52 weeks).
Any new result cancels out the corresponding one from the previous year, if present, so the rankings are updated approximately every week \cite{ATP}.
Different tournaments reward players with different point scales.
Among the competitions we considered in this work, Challengers are the less prestigious, as players can be rewarded from 3 up to 125 points, whereas Grand Slams are the most prestigious, as players' rewards range from 8 to 2000 points, depending on which round they reached.
The other tourneys fall in between: Masters 1000 points vary from 8 to 1000; ATP 500 points scale from 4 to 500; ATP 250 points range from 3 to 250.
Detailed scales per tournament are available in the SI (Table~S2).

\subsection*{Statistics of match wins}
In \cref{fig:3}A, we show that players belonging to a group $i$ have a probability $P_i(T \geq t)$ of attending at least $t$ tourneys where they win a match, within the first ten tournaments of their career in the ATP.
This probability results from the cumulative distribution of the function $p_i(t)$:
\begin{equation}
P_i(T \geq t) = \int_{t}^{\infty} p_i(t')dt'
\label{eq:1}
\end{equation}
Where $p_i(t)$ is the fraction of players of a group $i = \{ \mathrm{top, middle, bottom} \}$ with a win $w$ in exactly $t$ attended tourneys, with $1 \leq t \leq 10$, namely:
\begin{equation}
p_i(t) = \frac{N_i(w=t)}{\sum_{s=1}^{10}N_i(w=s)}
\label{eq:2}
\end{equation}

We also consider the probability $P$ that players have won their first match since becoming professional in a given tourney $t$ (\cref{fig:3}B).
\cref{eq:3} reports the fraction of players who have their first win at time $t$, that is, at the tournament $1 \leq t \leq 100$, grouped by their career peak.
Given the players in the group $i$ with their first win $w^*$ at the time $t$, we can write the following:
\begin{equation}
P_i(t,w^*) = \frac{N_i(t,w^*)}{N_i(t)}
\label{eq:3}
\end{equation}
Where $P_i(t,w^*)$ is the fraction of players with their first win $w^*$ at time $t$, that is the ratio of the number of athletes $N_i(t,w^*)$ with their first win $w^*$ at time $t$, divided by the number of players who competed at time $t$, $N_i(t)$.

\subsection*{Network centrality}
The co-attendance network of tennis tournaments is based on the career trajectories of the players in our data.
This results in a weighted directed network, where nodes are tourneys and links $\left(i, j \right)$ are created when players first attend tournament $i$, then $j$.
Link weights are obtained by the number of times different players generate the same link.
Specifically, every link $\left(i, j \right)$ has a weight $\Tilde{\omega}_{ij} = \frac{\omega_{ij}}{\omega_{\mathrm{max}}}$, normalized to the maximum possible weight found in the network, i.e., $\omega_{\mathrm{max}}= \mathrm{max} \left( \omega_{ij} \right)$.
We use the topology of the co-attendance network to assess the prestige of tourneys.
Specifically, we rely on the eigenvector centrality $x_{i}$ \cite{bonacich1987power}, defined for a node $i$ in a directed network as proportional to the centralities of the nodes that point to $i$ \cite{newman2018networks}:
\begin{equation}
x_{i} = \kappa_{1}^{-1} \sum_{j} A_{ij} x_{j}
\label{eq:4}
\end{equation}
Where the term $\kappa_{1}$ represents the largest eigenvalue of the adjacency matrix $\mathbf{A}$ whose elements are $A_{ij}$.

\section*{Authors' contributions}
C.~Z. and T.~C. conceived the research.
C.~Z. performed the analysis and wrote the first draft of the manuscript. 
S.~S. provided methodological insights.
R.~S. provided the interpretation of some results and the null model formulation. 
T.~C., S.~S., and R.~S. supervised the study.
A.~P. and A.~R. helped supervise the project.
All authors discussed the results and commented on the manuscript.

\section*{Acknowledgments}
C.~Z. would like to thank L.~Gallo for valuable discussions and comments.
C.~Z. also acknowledges the support of 101086712-LearnData-HORIZON-WIDERA-2022-TALENTS-01 financed by EUROPEAN RESEARCH EXECUTIVE AGENCY (REA) (\url{https://cordis.europa.eu/project/id/101086712}), the Erasmus Mobility Network, and the Danish Data Science Academy (DDSA) for funding her visits to the research group of R.~S.
A.~P. and A.~R. acknowledge partial financial support of PRIN 2017WZFTZP \emph{Stochastic forecasting in complex systems}.
R.~S. and S.~S. acknowledge support from Villum Fonden through the Villum Young Investigator program (project number: 00037394).

\bibliography{library}

\begin{thebibliography}{10}

\bibitem{Sinatra2016}
Roberta Sinatra, Dashun Wang, Pierre Deville, Chaoming Song, and
  Albert~L{\'{a}}szl{\'{o}} Barab{\'{a}}si.
\newblock {Quantifying the evolution of individual scientific impact}.
\newblock {\em Science}, 354(6312), 2016.

\bibitem{Clauset2017}
Aaron Clauset, Daniel~B. Larremore, and Roberta Sinatra.
\newblock Data-driven predictions in the science of science.
\newblock {\em Science}, 355(6324):477--480, 2017.

\bibitem{Fortunato2018}
Santo Fortunato, Carl~T. Bergstrom, Katy Börner, James~A. Evans, Dirk Helbing,
  Staša Milojević, Alexander~M. Petersen, Filippo Radicchi, Roberta Sinatra,
  Brian Uzzi, Alessandro Vespignani, Ludo Waltman, Dashun Wang, and
  Albert-László Barabási.
\newblock Science of science.
\newblock {\em Science}, 359(6379):eaao0185, 2018.

\bibitem{bol2018matthew}
Thijs Bol, Mathijs De~Vaan, and Arnout van~de Rijt.
\newblock The matthew effect in science funding.
\newblock {\em Proceedings of the National Academy of Sciences},
  115(19):4887--4890, 2018.

\bibitem{bonaventura_predicting_2020}
Moreno Bonaventura, Valerio Ciotti, Pietro Panzarasa, Silvia Liverani, Lucas
  Lacasa, and Vito Latora.
\newblock Predicting success in the worldwide start-up network.
\newblock {\em Scientific Reports}, 10(1):345, January 2020.

\bibitem{Williams2019}
Oliver~E Williams, Lucas Lacasa, and Vito Latora.
\newblock Quantifying and predicting success in show business.
\newblock {\em Nature communications}, 10(1):1--8, 2019.

\bibitem{Fraiberger2018}
Samuel~P. Fraiberger, Roberta Sinatra, Magnus Resch, Christoph Riedl, and
  Albert~L{\'{a}}szl{\'{o}} Barab{\'{a}}si.
\newblock {Quantifying reputation and success in art}.
\newblock {\em Science}, 362(6416):825--829, 2018.

\bibitem{nadini2021mapping}
Matthieu Nadini, Laura Alessandretti, Flavio Di~Giacinto, Mauro Martino,
  Luca~Maria Aiello, and Andrea Baronchelli.
\newblock Mapping the nft revolution: market trends, trade networks, and visual
  features.
\newblock {\em Scientific reports}, 11(1):20902, 2021.

\bibitem{vasan2022quantifying}
Kishore Vasan, Mil{\'a}n Janosov, and Albert-L{\'a}szl{\'o} Barab{\'a}si.
\newblock Quantifying nft-driven networks in crypto art.
\newblock {\em Scientific reports}, 12(1):2769, 2022.

\bibitem{Salganik2006}
Matthew~J. Salganik, Peter~Sheridan Dodds, and Duncan~J. Watts.
\newblock Experimental study of inequality and unpredictability in an
  artificial cultural market.
\newblock {\em Science}, 311(5762):854--856, 2006.

\bibitem{janosov_elites_2020}
Milán Janosov, Federico Musciotto, Federico Battiston, and Gerardo Iñiguez.
\newblock Elites, communities and the limited benefits of mentorship in
  electronic music.
\newblock {\em Scientific Reports}, 10(1):3136, February 2020.

\bibitem{kang2022analyzing}
Inwon Kang, Michael Mandulak, and Boleslaw~K. Szymanski.
\newblock Analyzing and predicting success of professional musicians.
\newblock {\em Scientific Reports}, 12(1):21838, December 2022.

\bibitem{wang_success_2019}
Xindi Wang, Burcu Yucesoy, Onur Varol, Tina Eliassi-Rad, and Albert-László
  Barabási.
\newblock Success in books: predicting book sales before publication.
\newblock {\em EPJ Data Science}, 8(1):31, October 2019.

\bibitem{Janosov2020}
Mil{\'{a}}n Janosov, Federico Battiston, and Roberta Sinatra.
\newblock {Success and luck in creative careers}.
\newblock {\em EPJ Data Science}, 9(1), 2020.

\bibitem{murray_human_2003}
Charles~A. Murray.
\newblock {\em Human accomplishment: the pursuit of excellence in the arts and
  sciences, 800 {BC} to 1950}.
\newblock HarperCollins, New York, 2003.

\bibitem{Yucesoy2016}
Burcu Yucesoy and Albert~L{\'{a}}szl{\'{o}} Barab{\'{a}}si.
\newblock {Untangling performance from success}.
\newblock {\em EPJ Data Science}, 5(1), 2016.

\bibitem{lehmann2006measures}
Sune Lehmann, Andrew~D Jackson, and Benny~E Lautrup.
\newblock Measures for measures.
\newblock {\em Nature}, 444(7122):1003--1004, 2006.

\bibitem{Barabasi2018}
Albert~L{\'{a}}szl{\'{o}} Barab{\'{a}}si.
\newblock {\em The Formula: The Five Laws Behind Why People Succeed}.
\newblock Pan Macmillan, 2018.

\bibitem{Radicchi2012}
Filippo Radicchi.
\newblock {Universality, limits and predictability of gold-medal performances
  at the olympic games}.
\newblock {\em PLoS ONE}, 7(7), 2012.

\bibitem{Radicchi2011a}
Filippo Radicchi.
\newblock {Who is the best player ever? a complex network analysis of the
  history of professional tennis}.
\newblock {\em PLoS ONE}, 6(2), 2011.

\bibitem{tennant_complexity_2017}
Adam~G Tennant, Nasir Ahmad, and Sybil Derrible.
\newblock Complexity analysis in the sport of boxing.
\newblock {\em Journal of Complex Networks}, 5(6):953--963, December 2017.

\bibitem{pappalardo2018quantifying}
Luca Pappalardo and Paolo Cintia.
\newblock Quantifying the relation between performance and success in soccer.
\newblock {\em Advances in Complex Systems}, 21(03n04):1750014, May 2018.

\bibitem{Sobkowicz2020}
Pawel Sobkowicz, Robert~H. Frank, Alessio~E. Biondo, Alessandro Pluchino, and
  Andrea Rapisarda.
\newblock {Inequalities, chance and success in sport competitions: Simulations
  vs empirical data}.
\newblock {\em Physica A: Statistical Mechanics and its Applications},
  557:124899, 2020.

\bibitem{Zappala2022}
Chiara Zappalà, Alessandro Pluchino, Andrea Rapisarda, Alessio~Emanuele
  Biondo, and Pawel Sobkowicz.
\newblock On the role of chance in fencing tournaments: An agent-based
  approach.
\newblock {\em PLOS ONE}, 17(5):1--17, 05 2022.

\bibitem{Petersen2011}
Alexander~M. Petersen, Woo~Sung Jung, Jae~Suk Yang, and H.~Eugene Stanley.
\newblock {Quantitative and empirical demonstration of the Matthew effect in a
  study of career longevity}.
\newblock {\em Proceedings of the National Academy of Sciences of the United
  States of America}, 108(1):18--23, 2011.

\bibitem{zappala_paradox_2023}
Chiara Zappalà, Alessio~Emanuele Biondo, Alessandro Pluchino, and Andrea
  Rapisarda.
\newblock The paradox of talent: How chance affects success in tennis
  tournaments.
\newblock {\em Chaos, Solitons \& Fractals}, 176:114088, 2023.

\bibitem{mauboussin2012success}
Michael~J Mauboussin.
\newblock {\em The success equation: Untangling skill and luck in business,
  sports, and investing}.
\newblock Harvard Business Review Press, 2012.

\bibitem{Frank2016}
Robert~H. Frank.
\newblock {\em Success and luck: Good Fortune and the Myth of Meritocracy}.
\newblock Princeton University Press, Princeton, 2016.

\bibitem{perc2014matthew}
Matja{\v{z}} Perc.
\newblock The matthew effect in empirical data.
\newblock {\em Journal of The Royal Society Interface}, 11(98):20140378, 2014.

\bibitem{ATP}
Official site of men's professional tennis \uppercase{ATP} tour.
\newblock \url{https://www.atptour.com/}.

\bibitem{ATPgit}
Tennis data repository.
\newblock \url{https://github.com/JeffSackmann/tennis_atp}.

\bibitem{Breznik2015}
Kristijan Breznik.
\newblock {Revealing the best doubles teams and players in tennis history}.
\newblock {\em International Journal of Performance Analysis in Sport},
  15(3):1213--1226, 2015.

\bibitem{Aparicio2016}
David Apar{\'{i}}cio, Pedro Ribeiro, and Fernando Silva.
\newblock {A subgraph-based ranking system for professional tennis players}.
\newblock {\em Studies in Computational Intelligence}, 644:159--171, 2016.

\bibitem{bonacich1987power}
Phillip Bonacich.
\newblock Power and centrality: A family of measures.
\newblock {\em American journal of sociology}, 92(5):1170--1182, 1987.

\bibitem{wu1988estimation}
Margaret~C Wu and Raymond~J Carroll.
\newblock Estimation and comparison of changes in the presence of informative
  right censoring by modeling the censoring process.
\newblock {\em Biometrics}, pages 175--188, 1988.

\bibitem{Schulz1988}
Roy Schulz and Craig~S. Curnow.
\newblock Peak performance and age among superathletes: track and field,
  swimming, baseball, tennis, and golf.
\newblock {\em Journal of gerontology}, 43 5:P113--20, 1988.

\bibitem{Guillaume2011}
Marion Guillaume, Stephane Len, Muriel Tafflet, Laurent Quinquis, Bernard
  Montalvan, Karine Schaal, Hala Nassif, Fran{\c{c}}ois~Denis Desgorces, and
  Jean-Fran{\c{c}}ois Toussaint.
\newblock {Success and Decline}.
\newblock {\em Medicine {\&} Science in Sports {\&} Exercise},
  43(11):2148--2154, nov 2011.

\bibitem{Davidson-Pilon2019}
Cameron Davidson-Pilon.
\newblock lifelines: survival analysis in python.
\newblock {\em Journal of Open Source Software}, 4(40):1317, 2019.

\bibitem{Baker2013}
Joseph Baker, Dan Koz, Ann-Marie Kungl, Jessica Fraser-Thomas, and Jörg
  Schorer.
\newblock Staying at the top: playing position and performance affect career
  length in professional sport.
\newblock {\em High Ability Studies}, 24(1):63--76, 2013.

\bibitem{newman2018networks}
Mark Newman.
\newblock {\em Networks}.
\newblock Oxford university press, 2018.

\bibitem{Reid2007}
Machar Reid, Miguel Crespo, Luca Santilli, Dave Miley, and James Dimmock.
\newblock The importance of the international tennis federation's junior boys'
  circuit in the development of professional tennis players.
\newblock {\em Journal of Sports Sciences}, 25(6):667--672, 2007.
\newblock PMID: 17454534.

\bibitem{Kovalchik2017}
Stephanie~A Kovalchik, Michael~K Bane, and Machar Reid.
\newblock {Getting to the top: an analysis of 25 years of career rankings
  trajectories for professional women's tennis}.
\newblock {\em Journal of Sports Sciences}, 35(19):1904--1910, oct 2017.

\bibitem{Pingwei2018}
Pingwei Li, Veerle~De Bosscher, and Juanita~R. Weissensteiner.
\newblock The journey to elite success: a thirty-year longitudinal study of the
  career trajectories of top professional tennis players.
\newblock {\em International Journal of Performance Analysis in Sport},
  18(6):961--972, 2018.

\bibitem{Brouwers2012}
Jessie Brouwers, Veerle {De Bosscher}, and Popi Sotiriadou.
\newblock An examination of the importance of performances in youth and junior
  competition as an indicator of later success in tennis.
\newblock {\em Sport Management Review}, 15(4):461--475, 2012.

\end{thebibliography}


\begin{thebibliography}{1}

\bibitem{scott2015multivariate}
David~W Scott.
\newblock {\em Multivariate density estimation: theory, practice, and
  visualization}.
\newblock John Wiley \& Sons, 2015.

\bibitem{Guillaume2011}
Marion Guillaume, Stephane Len, Muriel Tafflet, Laurent Quinquis, Bernard
  Montalvan, Karine Schaal, Hala Nassif, Fran{\c{c}}ois~Denis Desgorces, and
  Jean-Fran{\c{c}}ois Toussaint.
\newblock {Success and Decline}.
\newblock {\em Medicine {\&} Science in Sports {\&} Exercise},
  43(11):2148--2154, nov 2011.

\bibitem{saramaki2007generalizations}
Jari Saram{\"a}ki, Mikko Kivel{\"a}, Jukka-Pekka Onnela, Kimmo Kaski, and Janos
  Kertesz.
\newblock Generalizations of the clustering coefficient to weighted complex
  networks.
\newblock {\em Physical Review E}, 75(2):027105, 2007.

\bibitem{bonacich1987power}
Phillip Bonacich.
\newblock Power and centrality: A family of measures.
\newblock {\em American journal of sociology}, 92(5):1170--1182, 1987.

\bibitem{newman2018networks}
Mark Newman.
\newblock {\em Networks}.
\newblock Oxford university press, 2018.

\bibitem{ATP}
Official site of men's professional tennis - \uppercase{ATP} tour.
\newblock \url{https://www.atptour.com/}.

\end{thebibliography}

\end{document}


\flushbottom
\maketitle
\thispagestyle{empty}

\section*{Supplementary Information}
\subsection*{The effect of active players}
\begin{figure}[h!]
\centering
\includegraphics[width=\textwidth]{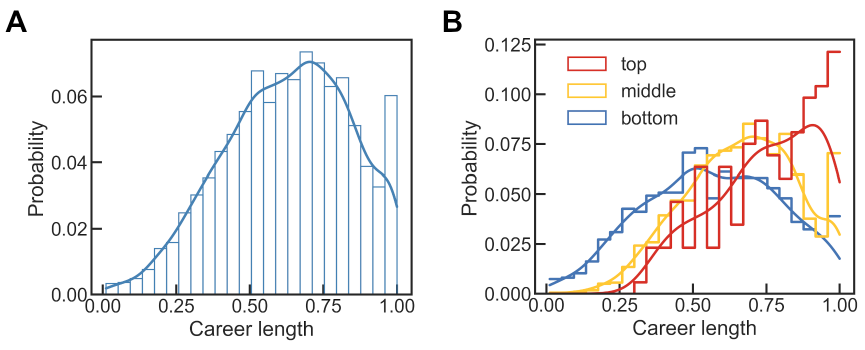}
\caption{Career peak distributions, active players included.
\textbf{A} Distribution for the whole community of players.
\textbf{B} Distribution after splitting players in groups. 
Top players (in red) have different behavior, most likely due to their longer career at the professional level.
Histograms are normalized so that bar heights sum to 1, and are reported with a Kernel Density Estimation of the data (continuous curves).}
\label{fig:SI1}
\end{figure}

The ATP dataset we analyzed consists of players who have appeared for at least two years in the official ranking.
We only consider players that started their careers within our dataset, so from 2000 on.
However, some of those players could still be active at the end of 2019, which is the upper bound of our dataset.
In the main text, we controlled for right-censored data, which occur when the time of observation ends before a certain event, when we examined the time of the peak along a career, in terms of ranking points.
In panels B-C of Fig.~1, we considered a subset of 2,262 players who started and ended their careers within our observation time.
Here, we show the time of the career peak if we include active players (\cref{fig:SI1}).
In this case, we see that top players (red line) are the only ones with different behavior, most likely because their careers tend to last longer; thus, they either have not reached their peak yet or their amount of points is stable.

We specify that players might reach their maximum number of points more than once in their professional career.
We tackle the possibility of multiple peaks by choosing the time of their peak at random.
It is worth mentioning that the continuous curves of \cref{fig:SI1} derive from a Kernel Density Estimation of the data \cite{scott2015multivariate}.

\begin{figure}[h!]
\centering
\includegraphics[scale=.65]{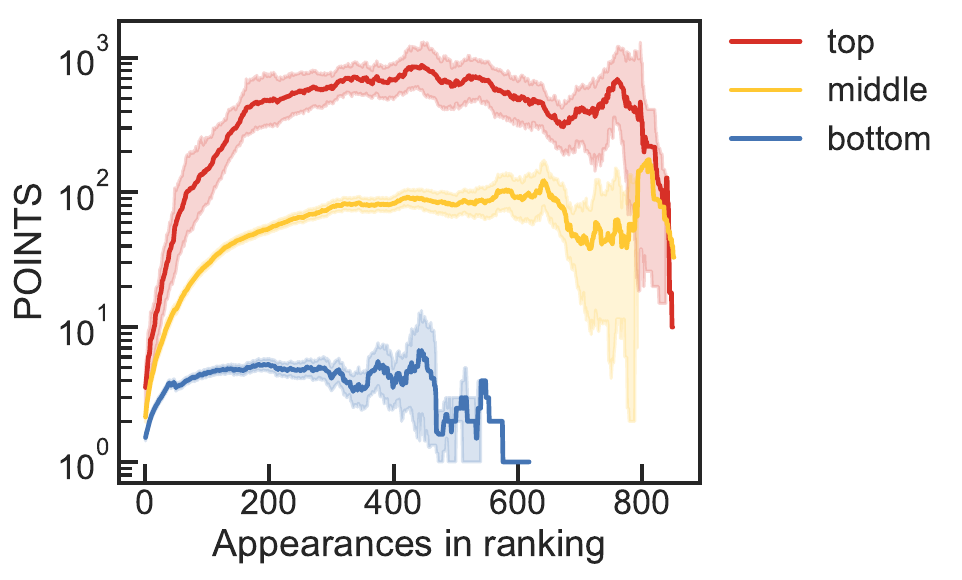}
\caption{Average trend of male tennis players in ATP ranking, active players excluded.
The tail of the evolution of points for the top players (red curve) suddenly drops, while the decline is smoother for the middle (yellow) and bottom (blue) players.}
\label{fig:SI2}
\end{figure}

The presence of active players in the data also impacts the evolution of their ranking points over time.
Indeed, the curves in panel B of Fig.~2 all seem to decline as they approach the end of the observation time.
This could suggest that players experience an overall decrease in their points before the end of their careers due to a reduction in the number of tournaments played or an increase in poorer performances \cite{Guillaume2011}.
However, we can appreciate the decline in the tail of these timelines only if we disentangle the contribution of active players.
Therefore, in \cref{fig:SI2}, we show the average trend of points in terms of ranking appearances for players who started and ended their careers within the dataset.
We observe a steep decrease in the ranking points of the top players, more evident than in the other groups.

\subsection*{Network features}
As explained in the main text, we built the co-attendance network of tennis tournaments based on the trajectories of players along their careers.
This results in a weighted directed network, where nodes are tourneys and links $\left(i, j \right)$ are created when players first attend tournament $i$, then $j$.
In \cref{tab:net_data} we summarize the main characteristics of this network, which is a dense and highly clustered graph.
These characteristics could come from the seasonality of the ATP tour, which ``forces'' athletes to repeat their trajectories to preserve their ranking points from the previous year, if not improve them.  
\begin{table}[h!]
\centering
\begin{tabular}{lc}
\hline
Feature & Value \\
\hline
Nodes & 651 \\
Links & 254583 \\
Density & 0.60\\
Clustering & 0.77\\
\hline
\end{tabular}
\caption{Summary of the features of the co-attendance network.}
\label{tab:net_data}
\end{table}

In detail, the density of the network can be calculated as the ratio:
\begin{equation}
\frac{m}{n(n-1)}
\label{eq:dens}
\end{equation}
Where $m$ is the number of links and $n$ the number of nodes in the network.
The average clustering coefficient is defined as follows:
\begin{equation}
C = \frac{1}{n}\sum_i \frac{2t_i}{k_i\left(k_i - 1 \right)}
\label{eq:clust}
\end{equation}
Where $n$ is the number of nodes, $k_i$ is the degree of node $i$ and $t_i$ the number of triangles having node $i$ as one of the vertices \cite{saramaki2007generalizations}.
For simplicity, only in the case of \cref{eq:clust} we assume an undirected and unweighted network.

\begin{figure}[h!]
\centering
\includegraphics[scale=.65]{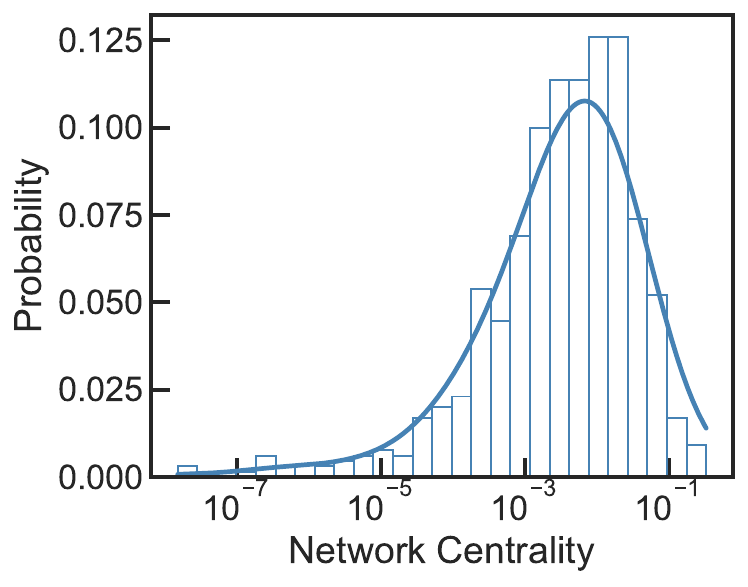}
\caption{Distribution of the eigenvector centrality for the tourneys (i.e., the nodes) of the co-attendance network.
It is asymmetric around its peak.
Bar heights sum to 1, the curve shows a Kernel Density Estimation of the data.}
\label{fig:SI3}
\end{figure}

We use the topology of the co-attendance network to assess the prestige of tourneys.
Specifically, we rely on the eigenvector centrality \cite{bonacich1987power,newman2018networks}.
For this network, the distribution of the eigenvector centrality is asymmetric, as shown in \cref{fig:SI3}.
\subsection*{Tournaments and their impact on players' careers}
The centrality of the competitions in the network is in agreement with the historical level of the tournaments (see Fig.~4B), expressed by the differences in reward for each type of tourney \cite{ATP}, which are summarized below (\cref{tab:tennis_points_chal}).
It is worth mentioning that here we refer to the ATP rulebook of 2019.
Since then, newer versions of the rulebook have been proposed.
\begin{table}[h!]
\centering
\begin{tabular}{lccccc}
\hline
 & Grand Slam & Masters 1000 & ATP 500 & ATP 250 & Challenger \\
\hline
Winner & 2000 & 1000 & 500 & 250 & 125\\
Final & 1200 & 600 & 300 & 150 & 75\\
Semifinal & 720 & 360 & 180 & 90 & 45\\
Quarter-final & 360 & 180 & 90 & 45 & 25\\
Round-of-16 & 180 & 90 & 45 & 20 & 10\\
Round-of-32 & 90 & 45 & 20 & 10 & 5\\
Round-of-64 & 45 & 25 & - & - & -\\
Round-of-128 & 10 & 10 & - & - & -\\
Qualif-\nth{1} & 25 & 16 & 10 & 5 & -\\
Qualif-\nth{2} & 16 & 8 & 4 & 3 & -\\ 
Qualif-\nth{3} & 8 & - & - & - & -\\
\hline
\end{tabular}
\caption{Allocation of points per tournament and round.
Points are assigned to the losers of the indicated round.
Please note that draws do not have a fixed length (except for Grand Slams, which always have a 128-draw).
Here, we show the points players can receive in tournaments with the maximum possible draw per type.
}
\label{tab:tennis_points_chal}
\end{table}

\begin{figure}[h!]
\centering
\includegraphics[scale=.65]{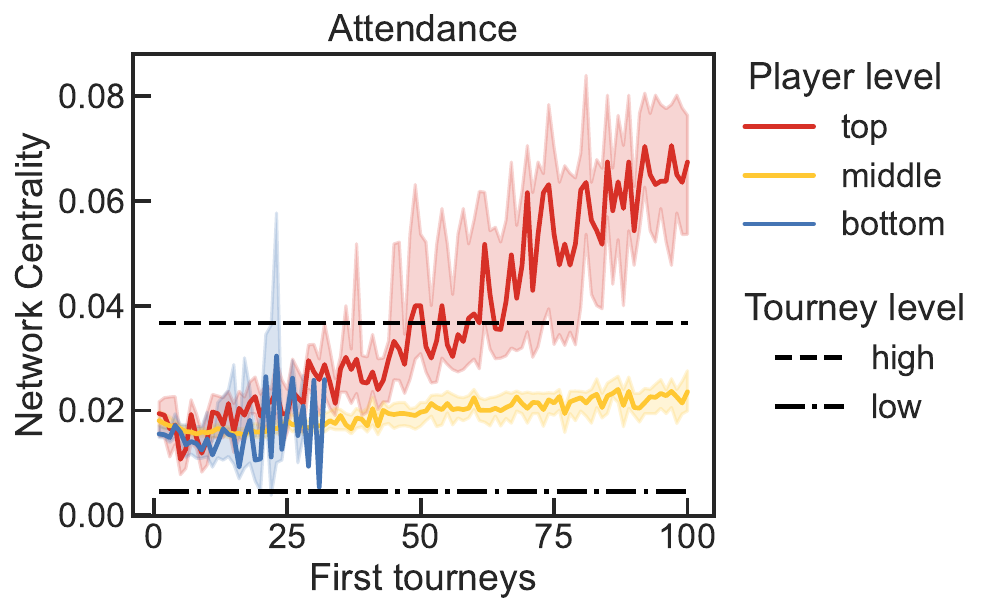}
\caption{Centrality for the first 100 attended tournaments per group of players.
The solid lines show the median trend and the shadows represent their confidence interval.
The dashed (dashed-dotted) lines refer to the high (low) level threshold of tourney splitting.
Top players (red dots) consistently compete in high-level tournaments only after 60 partipations.}
\label{fig:SI4}
\end{figure}

In the main text, we showed that the centrality of the first ten tournaments players attend is not associated with players' future success.
As we can see in \cref{fig:SI4}, the top players can be distinguished from the others by the prestige of tourneys attended only after 40 competitions, and they consistently compete in high-level venues around the \nth{60} tournament.

\begin{figure}[h!]
\centering
\includegraphics[width=\textwidth]{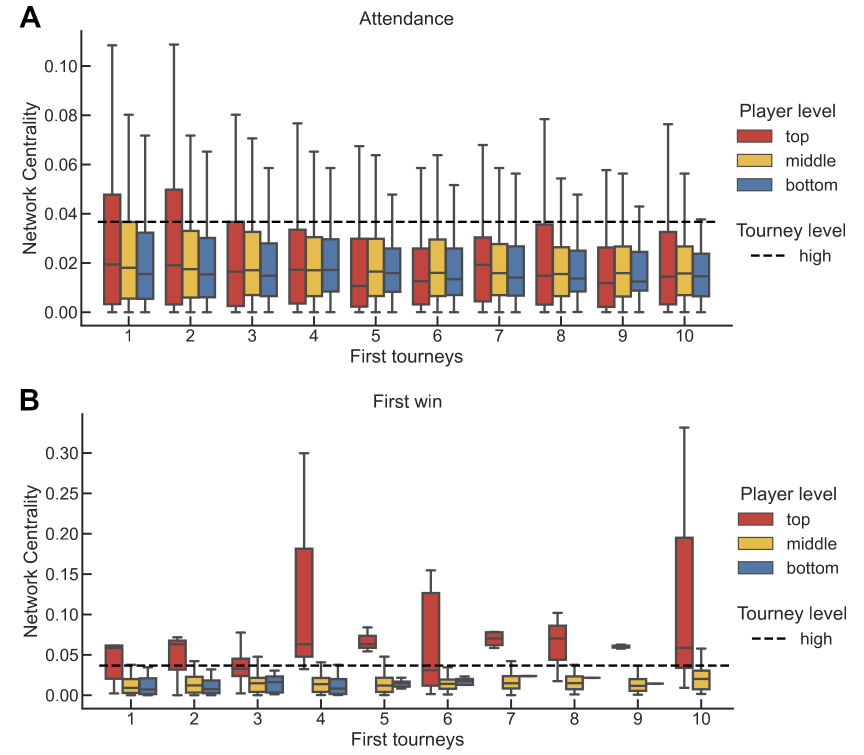}
\caption{Centrality distribution for each of the first ten tournaments, divided by player level.
\textbf{A} Boxplots of the eigenvector centrality of the tourneys that players attend (grouped by their career peak).
The average level of tourneys is below the high centrality threshold (dashed line). 
\textbf{B} Boxplots of the eigenvector centrality referred to players' first match win.
Only top players cross the high-level threshold consistently (dashed line).}
\label{fig:SI5}
\end{figure}

In addition, we focus on the centrality distribution for each of the tournaments we consider (from tournament 1 to tournament 10 per player level).
Again, we observe a common trend among the groups of players, even in this fine-grained visualization (see Fig.~5E of the main text for the aggregated version), when we only consider participation (\cref{fig:SI5}A).
In other words, there is no appreciable distinction in the average level of the first ten tourneys that players attend.
However, a difference emerges if we consider the level of the tournament in which the players won a match for the first time, as shown in \cref{fig:SI5}B: The centrality distributions are often above the high-level threshold for the top players (red boxplots), confirming the robustness of the results (see Fig.~5F of the main text for an aggregated visualization of the distributions).

\subsection*{Robustness of the impact of the first win}
\begin{figure}[p]
\centering
\includegraphics[width=\textwidth]{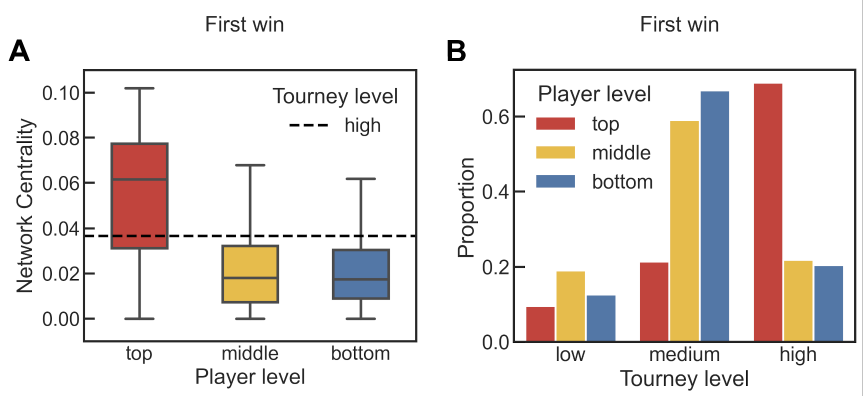}
\caption{First win with qualification rounds.
Both panels are based on the level of tournaments in which players have their first match win within the first ten attended competitions, and the top players (red) show distinct behavior compared to the others.
\textbf{A} Distribution of the centrality of the first win for each group of players.
The dashed line identify the threshold of high-level tourneys.
\textbf{B} Fractions of players who have their first win in a tournament of a given level.
Bars of the same color, each identifying a given group of players, add to 1.}
\label{fig:SI6}
\end{figure}

\begin{figure}[p]
\centering
\includegraphics[width=\textwidth]{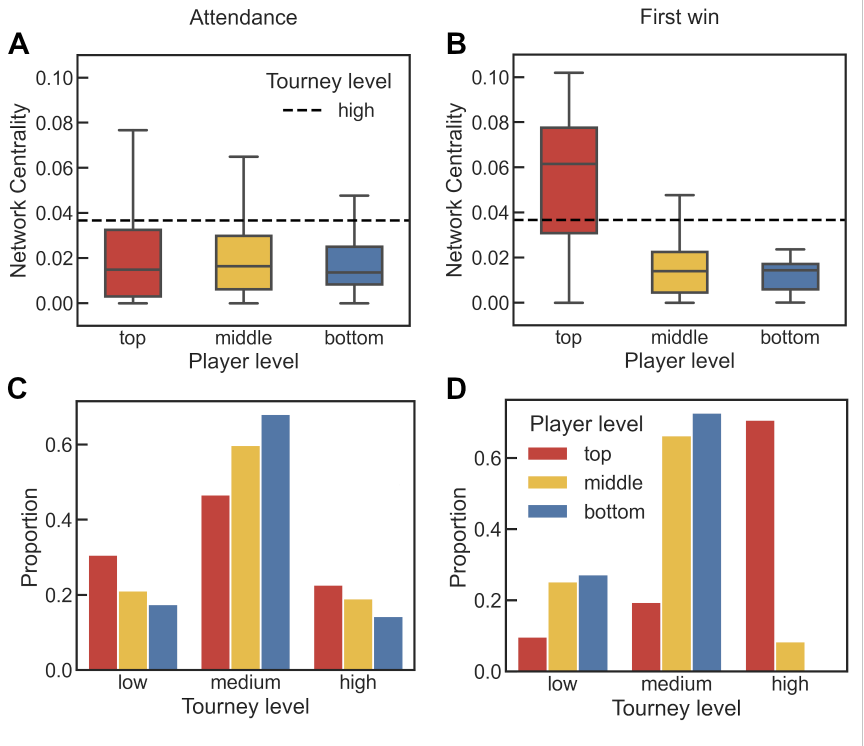}
\caption{Attendance and first win, considering players with more than ten tournaments in our dataset.
The panels on the left are based on the level of tourneys that the players attend, and no significant differences emerge among the groups.
The panels on the right are based on the level of tourneys where players have their first match win, and the top players show distinct behavior compared to the others.
\textbf{A} Distribution of the centrality of the first ten tournaments for each group of players.
\textbf{B} Distribution of the centrality of the first win for each group of players.
In panels A-B, the dashed line identify the threshold of high-level tourneys.
\textbf{C} Fractions of players who participated in a tournament of a given level within the first ten competitions.
\textbf{D} Fractions of players who have their first win in a tournament of a given level within their initial ten competitions.
In panels C-D, bars of the same color, each identifying a given group of players, add to 1.}
\label{fig:SI7}
\end{figure}

\begin{figure}[p]
\centering
\includegraphics[width=\textwidth]{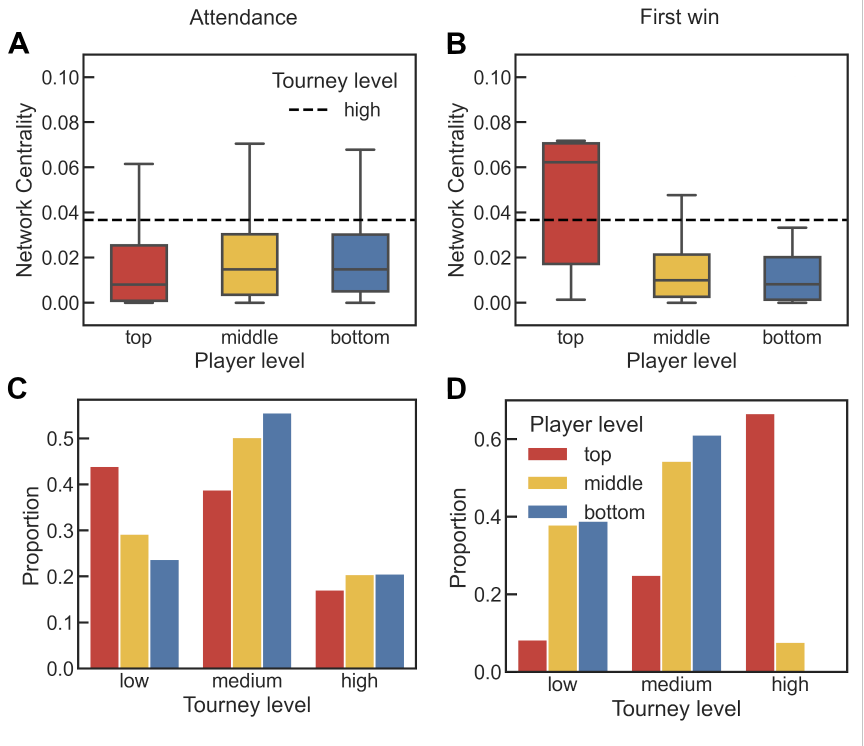}
\caption{Attendance and first win, excluding players that were still active at the end of our dataset.
The panels on the left are based on the level of tourneys that the players attend, and no significant differences emerge among the groups.
The panels on the right are based on the level of tourneys where players have their first match win, and the top players show distinct behavior compared to the others.
\textbf{A} Distribution of the centrality of the first ten tournaments for each group of players.
\textbf{B} Distribution of the centrality of the first win for each group of players.
In panels A-B, the dashed line identify the threshold of high-level tourneys.
\textbf{C} Fractions of players who participated in a tournament of a given level within the first ten competitions.
\textbf{D} Fractions of players who have their first win in a tournament of a given level within their initial ten competitions.
In panels C-D, bars of the same color, each identifying a given group of players, add to 1.}
\label{fig:SI8}
\end{figure}

We check that the predictive power of the first win still holds if we add (or remove) some constraints.
In detail, we verified that the first win within the first ten tournaments of players at the beginning of their careers identifies the top players in the following cases: if we count the qualification rounds (\cref{fig:SI6}); if we consider only players that have more than competitions in our dataset (\cref{fig:SI7}); if we exclude active players (\cref{fig:SI8}).
In all those scenarios, we can still observe that the top players act differently from the middle/bottom players.
Note that in \cref{fig:SI6} we did not report the level of tournaments attended because it does not change, compared to the one shown in the main text (Figs.~5C-E).


\bibliography{suppl_biblio}